\documentclass{elsart}
\usepackage{epsfig}

\begin{document}
\begin{frontmatter}
\title{\large\bf\boldmath Search for the Rare Decays $J/\psi
\rightarrow D^-_s e^+ {\nu}_e$, $J/\psi \rightarrow D^- e^+
{\nu}_e$, and $J/\psi \rightarrow \overline{D^0} e^+ e^-$}

\begin{small}
\begin{center}
\vspace{0.3cm}

M.~Ablikim$^{1}$,              J.~Z.~Bai$^{1}$,
Y.~Ban$^{12}$, J.~G.~Bian$^{1}$,              X.~Cai$^{1}$,
H.~F.~Chen$^{17}$, H.~S.~Chen$^{1}$,
H.~X.~Chen$^{1}$,              J.~C.~Chen$^{1}$, Jin~Chen$^{1}$
% ...
\\
\vspace{0.2cm}
(BES Collaboration)\\
\vspace{0.2cm} {\it
$^{1}$ Institute of High Energy Physics, Beijing 100049, People's Republic of China\\
% ...
$^{12}$ Peking University, Beijing 100871, People's Republic of China\\
% ...
$^{17}$ University of Science and Technology of China, Hefei 230026, People's Republic of China\\
% ...
}

\end{center}
\end{small}
\maketitle

\normalsize

\begin{abstract}

We report on a search for the decays $J/\psi \rightarrow D^-_s e^+
{\nu}_e + c.c.$, $J/\psi \rightarrow D^- e^+ {\nu}_e + c.c.$, and
$J/\psi \rightarrow \overline{D^0} e^+ e^- + c.c.$ in a sample of
$5.8 \times 10^7 J/\psi$ events collected with the BESII detector
at the BEPC. No excess of signal above background is observed, and
90\% confidence level upper limits on the branching fractions are
set: ${B}(J/\psi \rightarrow D^-_s e^+ {\nu}_e +
c.c.)<4.8{\times}10^{-5}$, ${B}(J/\psi\rightarrow D^- e^+ {\nu}_e
+ c.c.)<1.2{\times}10^{-5}$, and ${B}(J/\psi \rightarrow
\overline{D^0} e^+ e^- + c.c.)<1.1{\times}10^{-5}$.

\vspace{3\parskip} \noindent{\it PACS:} 13.25.Gv, 13.30.Ce

\end{abstract}
\end{frontmatter}


\section{INTRODUCTION}
\label{introduction}

Hadronic, electromagnetic, and radiative decays of
the $J/\psi$ have been widely studied. However there have been few
searches for rare weak $J/\psi$ decay processes.  Kinematically, the
$J/\psi$ cannot decay into a pair of charmed $D$ mesons, but can decay into
a single $D$ meson. Searches for weak decays of $J/\psi$ to single
$D$ or $D_s$ mesons provide tests of standard model (SM) theory and
serve as a probe of new physics~\cite{zhangxm}, such as TopColor
models, the minimal supersymmetric standard model with or
without R-parity, and the two Higgs doublet model~\cite{topcol}.

\begin{figure}[htbp]
  \centering
  \begin{minipage}[b]{0.4\linewidth}
  \centering
  \includegraphics[height=3.0cm,width=4.5cm]{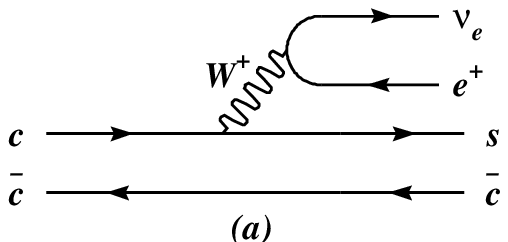}
  \end{minipage}
  \begin{minipage}[b]{0.4\linewidth}
  \centering
  \includegraphics[height=3.0cm,width=4.5cm]{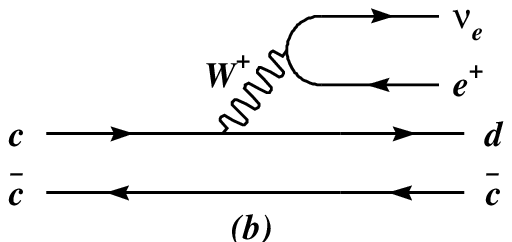}
  \end{minipage}
  \begin{minipage}[b]{0.4\linewidth}
  \centering
  \includegraphics[height=3.0cm,width=4.5cm]{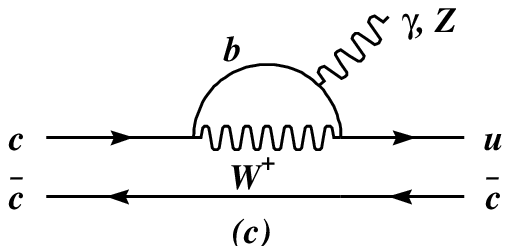}
  \end{minipage}
  \caption{
  Typical Feynman diagrams for (a) $J/\psi \rightarrow D^-_s e^+
  {\nu}_e$, (b) $J/\psi \rightarrow
  D^- e^+ {\nu}_e$, and (c) $J/\psi \rightarrow \overline{D^0} e^+ e^-$.}
  \label{feynme}
\end{figure}

The branching fractions of $J/\psi$ decays to single $D$ or $D_s$
mesons are predicted to be about $10^{-8}$ or smaller~\cite{zphy}
in the SM. The flavor changing neutral current (FCNC) process
$c\rightarrow{u}$ occurs in the standard model only at the loop
level where it is suppressed by the GIM mechanism.
Fig.~\ref{feynme} shows the dominant Feynman diagrams within the
standard model for the decays $J/\psi \rightarrow D^-_s e^+
{\nu}_e$, $J/\psi \rightarrow D^- e^+ {\nu}_e$, and $J/\psi
\rightarrow \overline{D^0} e^+ e^-$. No decay of this type has
been observed before. In this paper, we perform a search for the
decays $J/\psi \rightarrow D^-_s e^+ {\nu}_e$, $J/\psi \rightarrow
D^- e^+ {\nu}_e$, and $J/\psi \rightarrow \overline{D^0} e^+ e^-$
in a sample of $5.8{\times}10^7 J/\psi$ events collected with the
Beijing Spectrometer (BESII)~\cite{besii} detector at the Beijing
Electron-Positron Collider (BEPC)~\cite{bepc}. Throughout this
paper the charge conjugate states are implicitly included.

\section{BESII Detector}
\label{besii}

BES is a conventional solenoidal magnetic detector that is
described in detail in Ref.~\cite{bes}. BESII is the upgraded
version of the BES detector~\cite{besii}. A 12-layer Vertex
Chamber (VC) surrounds the beryllium beam pipe and provides track
and trigger information. A forty-layer main drift chamber (MDC)
located just outside the VC provides measurements of charged
particle trajectories over $85\%$ of the total solid angle; it
also provides ionization energy loss ($dE/dx$) measurements which
are used for particle identification (PID).  A momentum resolution
of $1.78\%\sqrt{1+p^2}$ ($p$ in GeV/$c$) and a $dE/dx$ resolution
for Bhabha electrons of $\sim$8\% are obtained.  An array of 48
scintillation counters surrounding the MDC measures the time of
flight (TOF) of charged particles with a resolution of about 200
ps for hadrons. Outside the TOF counters, a 12 radiation length,
lead-gas barrel shower counter (BSC), operating in self quenching
streamer mode, measures the energies and positions of electrons
and photons over $80\%$ of the total solid angle with resolutions
of $\sigma_{E}/E=0.21/\sqrt{E}$ ($E$ in GeV), $\sigma_{\phi}=7.9$
mrad, and $\sigma_{z}=2.3$ cm. Outside the solenoidal coil, which
provides a 0.4 T magnetic field over the tracking volume, is an
iron flux return that is instrumented with three double-layer muon
counters that identify muons with momentum greater than 500~
MeV$/c$.

In this analysis, a GEANT3-based Monte Carlo program (SIMBES) with
detailed consideration of the detector performance is used. The
consistency between data and Monte Carlo has been checked in many
physics channels from $J/\psi$ and $\psi(2S)$ decays, and the
agreement is reasonable, as described in detail in
Ref.~\cite{simbes}.\par

\section{Event Selection}
\label{evtsel}

Each charged track is required to be well fitted to a helix that is
within the polar angle region $|\cos \theta|<0.8$ and to originate
from the beam interaction region, which is defined to be within 2 cm
of the beam line in the transverse plane and within 20 cm of the
interaction point along the beam direction.  For particle
identification, confidence levels ($CL$) are calculated for each
particle hypothesis using combined time-of-flight~\cite{tof} and MDC
energy loss information.  Pions and kaons are identified by requiring
the confidence level for the desired hypothesis to be greater than
0.1\% and, further, by requiring the normalized weights, defined as
$CL_{\alpha}/(CL_{\pi}+CL_{K})$, where $\alpha$ denotes the desired
particle, to exceed $0.7$. For electron identification, the dE/dx,
TOF, and BSC information are combined to form a particle
identification confidence level for the electron hypothesis. An
electron candidate is required to have $CL_{e}>1.0\%$ and satisfy
$CL_{e}/(CL_{e}+CL_{\pi}+CL_{K})>0.85$.

An isolated neutral cluster is considered to be a photon candidate
when the angle between the nearest charged track and the cluster
is greater than 18$^{\circ}$, the first hit is in the beginning
six radiation lengths, the difference between the angle of the
cluster development direction in the BSC and the photon emission
direction is less than 37$^{\circ}$, and the energy deposit in the
shower counter is greater than 60~MeV.

\subsection{$J/\psi\rightarrow D^-_s e^+{\nu}_e$ and $J/\psi\rightarrow D^- e^+{\nu}_e$}
\label{dsenudpenu}

Both {$J/\psi\rightarrow D^-_s e^+{\nu}_e$ and $J/\psi\rightarrow D^-
e^+{\nu}_e$ candidate decays must contain an $e^+$. $D^{-}_{s}$ mesons are
reconstructed in two modes $D^{-}_{s}\rightarrow \phi {\pi}^-$ and
$K^{-}{K}^{*0}$, with $\phi \rightarrow{K}^+ K^-$ and
${K}^{*0}\rightarrow {K}^{+}{\pi}^{-}$.  The $\phi$ candidates are
reconstructed from two oppositely-charged kaons and must have an
invariant mass $|M_{KK}-1.02|<0.015$~GeV/$c^2$. The $K^{*0}$
candidates are constructed from $K^+$ and ${\pi}^{-}$ candidates and
are required to have an invariant mass in the range
$|M_{K\pi}-0.896|<0.060$~GeV/$c^2$.

$D^-$ candidates are reconstructed in the mode $D^- \rightarrow
K^+{\pi}^-{\pi}^-$.  Candidate events are required to have four
charged tracks which satisfy charged track selection criteria, and the
total charge of the tracks is required to be zero.

The polarization of the $K^{*0}$ meson in $D^{-}_{s}$ decay is
also utilized to reject backgrounds by imposing a selection
requirement on the helicity angle ${\theta}_{H}$. The helicity
angle is defined as the angle between one of the decay products of
the $K^{*0}$ and the direction of the flight of $K^{*0}$ in the
$K^{*0}$ rest frame. Background events are distributed uniformly
in $\cos\theta_{H}$ since they originate from random combinations,
while signal events are distributed as ${\cos}^{2}{\theta}_{H}$.
The $K^{*0}$ candidates are required to have
$\left|\cos\theta_{H}\right|>0.4$. Fig.~\ref{res} shows the
invariant mass distributions of $K^+K^-$ and $K^+\pi^-$ systems
with arrows indicating the selection of $\phi$ and $K^*(892)$
candidates.

\begin{figure}[htbp]
  \centering
  \includegraphics[width=7.0cm]{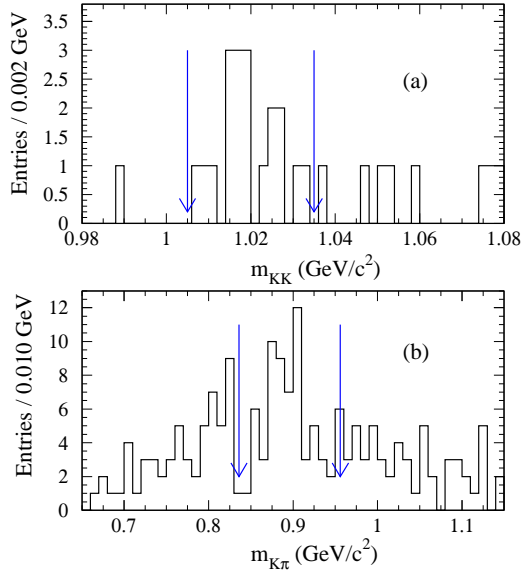}
  \caption{(a) $K^+K^-$ and (b) $K^+{\pi}$ invariant mass
  distributions of $D^{-}_{s}\rightarrow \phi {\pi}^-, ~\phi
  \rightarrow{K}^+ K^-$ and $D^{-}_{s}\rightarrow K^{-}{K}^{*0},
  ~{K}^{*0}\rightarrow {K}^{+}{\pi}^{-}$ candidates.  The signal
  regions for individual channels are indicated by arrows in the
  plot.}
  \label{res}
\end{figure}

The angle between the identified electron and the nearest charged
track is required to be larger than $12^{\circ}$ to reject
backgrounds from gamma conversions and $\pi^0$ Dalitz decays. In
order to reject backgrounds from $J/\psi$ decaying to states with
extra neutral particles, the number of isolated photons is
constrained to be zero. The requirements for an isolated photon
are more stringent than those for a good photon in order to
increase the detection efficiency.  An isolated photon is a photon
with the angle between the nearest charged track and the cluster
greater than 22$^{\circ}$ and the difference between the angle of
the cluster development direction in the BSC and the photon
emission direction less than 60$^{\circ}$, and the energy deposit
in the shower counter is greater than 0.1~GeV.

The four-momentum of the neutrino, $p_{\nu}=(E_{miss},~
\overrightarrow{p_{miss}})$, is inferred from the difference between
the net four momentum of the $J/\psi$ particle,
$p_{J/\psi}=(M_{J/\psi}, 0, 0, 0,)$, and the sum of the four-momentum
of all detected particles in the event. $P_{miss}$ is required to be
larger than $0.2$~GeV/$c$ and less than $0.9$~GeV/$c$ for $J/\psi
\rightarrow D^-_s e^+ {\nu}_e$ and larger than 0.2~GeV/$c$ and less
than $1.0$~GeV/$c$ for $J/\psi \rightarrow D^- e^+ {\nu}_e$ to reduce
background from $J/\psi$ decay to 4-prong final states with
misidentified particles but with no missing particles.  We further
require the absolute value of $U_{miss}$, which is defined as
$U_{miss}=E_{miss}-P_{miss}$, to be less than 0.1~GeV, to reject
backgrounds from $J/\psi$ decaying to $K^0_L$, $\eta$, and partially missing
$\pi^0$ final states, which were not rejected by prior criteria. After
all requirements, the invariant mass distributions of $D^{-}_{s} \to
K^+ K^- \pi^-$ and $D^- \to K^+ \pi^-\pi^-$ are shown in
Figs.~\ref{invmass}(a) and (b), respectively.

\begin{figure}[htbp]
  \centering
  \includegraphics[width=7.0cm]{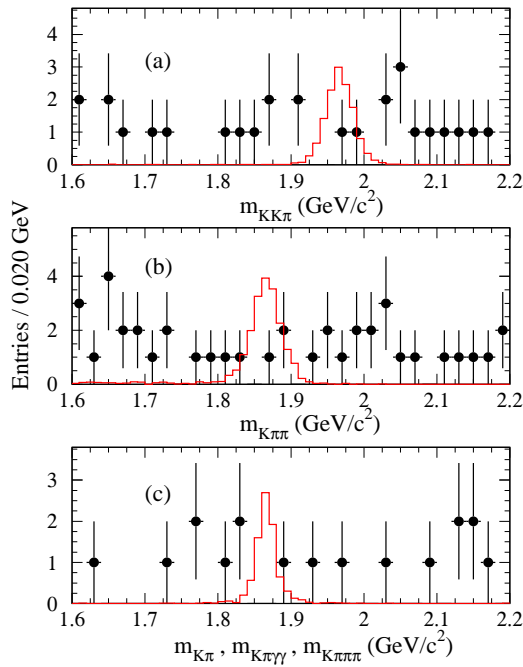}
  \caption{Invariant mass distribution of (a) $D^-_s\to K^+K^-\pi^-$,
  (b) $D^-\to K^+\pi^-\pi^-$, and (c)
  $\overline{D^0}\to {K}^+{\pi}^-,{K}^+{\pi}^-{\pi}^0,{K}^+{\pi}^-{\pi}^-{\pi}^+$. Data are shown by dots with error bars. The
  histogram is MC simulated signal events, unnormalized.}
  \label{invmass}
\end{figure}

\subsection{$J/\psi\rightarrow\overline{D^0} e^+ e^-$}
\label{d0ee}

$\overline{D^0}$ mesons are reconstructed in three decay modes:
$\overline{D^0}\rightarrow{K}^+{\pi}^-$,
$\overline{D^0}\rightarrow{K}^+{\pi}^-{\pi}^0$, and
$\overline{D^0}\rightarrow{K}^+{\pi}^-{\pi}^-{\pi}^+$. In
$\overline{D^0} \rightarrow {K}^+ {\pi}^{-} {\pi}^{0}$, the
${\pi}^{0}$ candidates are reconstructed from a pair of photons.
The $\pi^0$ decay angle $|E_{\gamma1}-E_{\gamma2}|/{P}_{\pi^0}$ is
required to be $<0.9$ for $K^-{\pi}^{+}{\pi}^{0}$ in order to
remove events in which the $\pi^0$ is falsely reconstructed from a
high-energy photon and a spurious shower. Here, $P_{\pi^0}$ is the
momentum of $\pi^0$ and $E_{\gamma1}$ and $E_{\gamma2}$ are the
energies of the two photons. Candidate events are required to have
four or six charged tracks which satisfy the charged track
selection criteria and total charge zero. Four-constraint (4C)
energy-momentum conservation kinematic fits are performed under
the $J/\psi \to e^+e^-{K}^{+}{\pi}^{-}$ or the $J/\psi \to
e^+e^-{K}^{+}{\pi}^{-}{\pi}^{-}{\pi}^{+}$ hypotheses, and the
${\chi}^2_{4C}$ is required to be less than 15. The $J/\psi\to
e^+e^-{K}^{+}{\pi}^{-}{\gamma}{\gamma}$ hypothesis is subjected to
a five-constraint (5C) fit where the invariant mass of the
${\gamma}{\gamma}$ pair associated with the ${\pi}^0$ is
constrained to $m_{{\pi}^0}$, and ${\chi}^2_{5C}<15$ is required.
For $\overline{D^0}\rightarrow{K}^{+}{\pi}^{-}{\pi}^{0}$, if more
than one combination of two good photons  passes the kinematic
fit, the combination with the smallest ${\chi}^2$ is chosen.
Gamma conversions and $\pi^0$ Dalitz decays may produce
electron-positron pairs with low electron-positron invariant mass. The
requirement $m_{ee}>0.2$~GeV/$c^2$ eliminates most of this type of
background.  The combined
$\overline{D^0}\to{K}^+{\pi}^-,{K}^+{\pi}^-{\pi}^0,{K}^+{\pi}^-{\pi}^-{\pi}^+$
invariant mass distribution is shown in Fig~\ref{invmass}(c).

\section{Monte Carlo Simulation}
\label{mcsim}

Monte Carlo (MC) simulation is used for the determination of mass
resolutions and detection efficiencies. We simulate
$J/\psi\rightarrow{D}^{-}_{s}e^+ {\nu}_e$, $J/\psi \rightarrow D^-
e^+ {\nu}_e$, and $J/\psi \rightarrow \overline{D^0} e^+ e^-$
production and decay, including the detector response. In these
simulations, 200000 events each are generated, and the decay
branching fractions of $D^-_s$, $D^-$, and $\overline{D^0}$ are
taken from the world average values~\cite{PDG}. The detection
efficiencies and branching fractions obtained are listed in Table
1.

\begin{table}[htbp]
  \centering
  \caption{Detection efficiencies and branching fractions~\cite{PDG}.}
  \begin{tabular}{|c|c|c|}
    \hline
    \hline
      Decay Mode & ${\varepsilon}_{i}$  & ${B}_{i}$  \\
    \hline
      $D^-_s\rightarrow \phi \pi^-$($\phi\rightarrow{K}^+K^-$) & 5.39\% & $(3.6{\pm}1.1)\%{\times}(49.1{\pm}0.6)\%$ \\
    \cline{2-3}
      $D^-_s\rightarrow K^- K^{*0}$($K^{*0}\rightarrow{K}^+{\pi}^-$) & 4.29\% & $(3.3{\pm}0.9)\%{\times}2/3$ \\
    \hline
      $D^-\rightarrow K^+ \pi^- \pi^-$ &  8.67\% & $(9.2{\pm}0.6)\%$ \\
    \hline
      $\overline{D^0}\rightarrow K^+ \pi^-$ &  5.24\% & $(3.80{\pm}0.09)\%$  \\
    \cline{2-3}
      $\overline{D^0}\rightarrow K^+ \pi^- \pi^0$ &  2.31\% & $(13.0{\pm}0.8)\%$ \\
    \cline{2-3}
      $\overline{D^0}\rightarrow K^+ \pi^- \pi^- \pi^+ $ &  2.15\% & $(7.46{\pm}0.31)\%$ \\
    \hline
    \hline
  \end{tabular}
\end{table}

\section{Systematic Errors}
\label{syserr}

The largest systematic errors on the branching fraction
measurements come from the uncertainties of the MDC simulation
(including systematic uncertainties of the tracking efficiency)
and the $U_{miss}$ or kinematic fit requirements. The photon
detection efficiency has been studied with several different
methods using $J/\psi\rightarrow{\rho}^0{\pi}^0$
decays~\cite{gam}; the difference between data and Monte Carlo
simulation is about 2\% for each photon. We estimate a systematic
error of 4\% for $\overline{D^0}\rightarrow{K}^+{\pi}^-{\pi}^0$.
The uncertainty for final states with no isolated photons is also
studied using $J/\psi\rightarrow{\rho}{\pi}$ decays and is about
2\%. The systematic error from electron identification is
estimated to be 5\% for each electron. The pion and kaon
identification is studied, and the difference between data and
Monte Carlo simulation is 1.5\% for each charged track, which is
treated as a systematic error. The error on the intermediate decay
branching fractions of $D^-_s$, $D^-$, $\overline{D^0}$, $\phi$,
and $K^{*0}$ are taken from the PDG~\cite{PDG}. The statistical
error of the Monte Carlo sample is taken into account. The total
number of $J/\psi$ events is
$(57.7{\pm}2.7){\times}10^{6}$~\cite{jpsinum}, determined from
inclusive 4-prong hadronic final states, and the uncertainty,
4.7\%, is taken as a systematic error. The systematic errors from
all sources, as well as the total, are listed in Table 2.

\begin{table}[htbp]
\scriptsize
  \centering
  \caption{Summary of the systematic errors.}
  \begin{tabular}{|c|c|c|c|c|c|}
    \hline
    \hline
      & {\scriptsize $J/\psi\rightarrow{D}^{-}_{s}e^+ {\nu}_e$} & {\scriptsize $J/\psi \rightarrow D^-
e^+ {\nu}_e$} & \multicolumn{3}{c|}{\scriptsize $J/\psi
\rightarrow \overline{D^0} e^+ e^-$} \\\hline
      & \scriptsize ${D}^{-}_{s}\rightarrow\phi \pi^-$, ${K}^-K^{*0}$ & \scriptsize ${D}^{-}\rightarrow{K}^+{\pi}^-{\pi}^-$ & \scriptsize $\overline{D^0}\rightarrow{K}^+ \pi^-$
      & \scriptsize ${K}^+ \pi^- \pi^0$ & \scriptsize ${K}^+ \pi^- \pi^- \pi^+$ \\
    \hline
     \scriptsize MDC Simulation & {18.6\%} & 11.6\% &
     \multicolumn{3}{c|}{20.6\%} \\
    \hline
     \scriptsize Gamma    & 2.0\% & 2.0\% & 0.0\% & 4.0\% & 0.0\% \\
    \hline
     \scriptsize $e$ PID    & 5.0\% & 5.0\% & \multicolumn{3}{c|}{10.0\%} \\
    \hline
     \scriptsize $\pi$, $K$ PID & 4.5\% & 4.5\% & 3.0\% & 3.0\% & 6.0\%\\
    \hline
     \scriptsize $B(D_s,D)$ & 25\% & 6.5\% & 2.4\% & 6.2\% & 4.2\% \\
    \hline
     \scriptsize MC Statistics & {5.1\%} & 2.5\% & \multicolumn{3}{c|}{2.7\%} \\
    \hline
     \scriptsize number of $J/\psi$ & \multicolumn{5}{c|}{4.7\%} \\
    \hline
     \scriptsize total    & 32.7\% & 15.9\% & 23.8\% & 24.8\% & 24.6\% \\
    \hline
    \hline
  \end{tabular}
\end{table}
\section{Results}
\label{results}

No excess of $J/\psi\rightarrow{D}^{-}_{s}e^+
{\nu}_e$, $J/\psi \rightarrow D^- e^+ {\nu}_e$, or $J/\psi \rightarrow
\overline{D^0} e^+ e^-$ events above background is observed. The upper
limit on the branching fractions of these decay modes are calculated
using
\begin{equation}
B<\frac{n^{obs}_{UL}}{N_{J/\psi}{\varepsilon}B(1-{\sigma}^{sys})},
\end{equation}
where $n^{obs}_{UL}$ is the upper limit of the observed number of
events at the 90\% confidence level. $N_{J/\psi}$ is the number of
$J/\psi$ events, and ${\varepsilon}B$ is defined as
${\sum\limits_{i=1}^{N}}{\varepsilon}_{i}B_i$, where
${\varepsilon}_{i}$ and $B_i$ are the detection efficiency and
branching fraction for decay channel $i$ and $N$ is the number of
decay modes, which are listed in Table 1. For instance, for
$J/\psi\rightarrow{D}^{-}_{s}e^+ {\nu}_e$, the $D^-_s$ mesons are
reconstructed in two decay modes, and $N$ is equal to 2. The
systematic error in the measurement is taken into consideration by
introducing $1-{\sigma}^{sys}$ in the denominator of the branching
fraction calculation.

We obtain upper limits for the observed number of events at 90\%
confidence level of 3.55 for $J/\psi\rightarrow{D}^{-}_{s} e^+
{\nu}_e$, 4.64 for $J/\psi \rightarrow D^{-} e^{+} {\nu}_{e}$ and
3.07 for $J/\psi\rightarrow \overline{D^0} e^+ e^-$, using a
Bayesian method with uniform prior above zero~\cite{PDG}. The
likelihood distributions and the 90\% C.L. limits are shown in
Fig.~\ref{uplim}. The likelihood values for each value of the
number of events are obtained by fitting the distributions shown
in Fig. 3 with a signal shape determined from MC simulation and a
second order polynomial to describe background. The upper limits
are obtained from the integral of the normalized likehood at the
90\% confidence level.  The numbers used in the branching fraction
calculations are summarized in Table 3.

\begin{table}[htbp]
\centering \caption{Numbers used in the calculation of upper limits
on the branching fractions of $J/\psi \rightarrow D^{-}_s/D^{-}
    e^{+} {\nu}_{e} + c.c.$ and $J/\psi\rightarrow \overline{D^0} e^+ e^- + c.c.$.} {\small
\begin{tabular}{|c|c|c|c|}
        \hline
        \hline
          & $J/\psi\rightarrow{D}^{-}_{s}e^+ {\nu}_e + c.c.$ & $J/\psi \rightarrow D^-
e^+ {\nu}_e + c.c.$ & $J/\psi \rightarrow \overline{D^0} e^+ e^- + c.c.$\\
        \hline
         $n^{obs}_{UL}$ & 3.55 & 4.64 & 3.07  \\
        \hline
         ${\varepsilon}{B}$ & $1.90{\times}10^{-3}$ & $7.98{\times}10^{-3}$ & $6.61{\times}10^{-3}$ \\
        \hline
         Sys. Err. & 32.7\% & 15.9\% & 24.8\% \\
        \hline
         $B$(90\%C.L.) & $<4.8{\times}10^{-5}$ &
         $<1.2{\times}10^{-5}$ & $<1.1{\times}10^{-5}$ \\
        \hline
        \hline
\end{tabular}
}
\end{table}

In summary, we have searched for the decays $J/\psi \rightarrow
D^-_s e^+ {\nu}_e$, $J/\psi \rightarrow D^- e^+ {\nu}_e$, and
$J/\psi \rightarrow \overline{D^0} e^+ e^-$ using $5.8 \times 10^7
J/\psi$ events acquired by the BESII detector at the BEPC $e^+
e^-$ collider. No evidence for any of these decays is found. The
final results for the 90\% confidence level upper limit of the
branching fractions are given in Table 3. The upper limits on the
branching fraction for decays $J/\psi \rightarrow D^-_s e^+
{\nu}_e$, $J/\psi \rightarrow D^- e^+ {\nu}_e$, and $J/\psi
\rightarrow \overline{D^0} e^+ e^-$ are not inconsistent with the
standard model.

\begin{figure}[htbp]
  \centering
  \includegraphics[width=7.0cm]{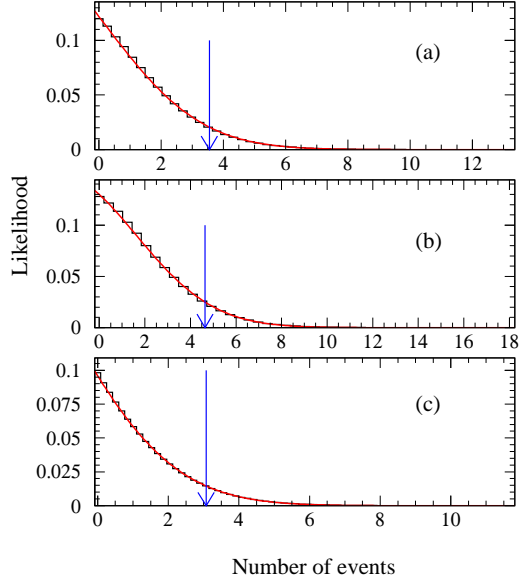}
  \caption{Likelihood distributions for the observed number of
  events of (a) $J/\psi\rightarrow{D}^{-}_{s}e^+ {\nu}_e$, (b)
  $J/\psi\rightarrow D^- e^+ {\nu}_e$, and (c) $J/\psi \rightarrow
  \overline{D^0} e^+ e^-$. The observed number of events at
  a Bayesian 90\% confidence level for individual channels are
  indicated by arrows in the plots.}
\label{uplim}
\end{figure}

The BES collaboration thanks the staff of BEPC and computing
center for their hard efforts. This work is supported in part by
the National Natural Science Foundation of China under contracts
Nos. 10491300, 10225524, 10225525, 10425523, the Chinese Academy
of Sciences under contract No. KJ 95T-03, the 100 Talents Program
of CAS under Contract Nos. U-11, U-24, U-25, and the Knowledge
Innovation Project of CAS under Contract Nos. U-602, U-34 (IHEP),
the National Natural Science Foundation of China under Contract
No. 10225522 (Tsinghua University), and the Department of Energy
under Contract No.DE-FG02-04ER41291 (U Hawaii).

\end{document}